\documentclass[twocolumn]{aastex701}

\usepackage{hyperref}
\usepackage{academicons}
\usepackage{gensymb}
\usepackage{graphicx}	
\usepackage{amsmath,amssymb}
\usepackage{xcolor}
\usepackage{appendix}
\usepackage{float}
\usepackage{tabularx}

\usepackage[T1]{fontenc}
\DeclareRobustCommand{\VAN}[3]{#2}
\let\VANthebibliography\thebibliography
\def\thebibliography{\DeclareRobustCommand{\VAN}[3]{##3}\VANthebibliography}

\begin{document}

\title{Presaging Doppler beaming discoveries of double white dwarfs during the Rubin LSST era}

\author[orcid=0000-0002-5864-1332,sname='Adamane Pallathadka']{Gautham Adamane Pallathadka}
\affiliation{William H. Miller III Department of Physics \& Astronomy, Johns Hopkins University, 3400 N Charles St, Baltimore, MD 21218, USA}
\email[show]{gadaman1@jh.edu}  

\author[orcid=0000-0002-0632-8897,sname=Zenati]{Yossef Zenati} 
\affiliation{Astrophysics Research Center of the Open University (ARCO), Department of Natural Sciences, Ra’anana 4353701, Israel}
\affiliation{William H. Miller III Department of Physics \& Astronomy, Johns Hopkins University, 3400 N Charles St, Baltimore, MD 21218, USA}
\email{yzenati1@jhu.edu}

\author[orcid=0000-0001-6100-6869,sname=Zakamska]{Nadia L. Zakamska}
\affiliation{William H. Miller III Department of Physics \& Astronomy, Johns Hopkins University, 3400 N Charles St, Baltimore, MD 21218, USA}
\affiliation{Institute for Advanced Study, Princeton, NJ 08540, USA}
\email{zakamska@jhu.edu}

\author[orcid=0000-0002-8820-4184,sname=Nguyen]{Ngan H. Nguyen}
\affiliation{William H. Miller III Department of Physics \& Astronomy, Johns Hopkins University, 3400 N Charles St, Baltimore, MD 21218, USA}
\email{nnguye53@jhu.edu}

\author[orcid=0000-0001-6806-0673,sname=Piro]{Anthony L.\ Piro}
\affiliation{The Observatories of the Carnegie Institution for Science, Pasadena, CA 91101, USA}
\email{piro@carnegiescience.edu}

\begin{abstract}
Double white dwarfs (DWDs) are by far the most common compact binaries in the Milky Way, are important low-frequency gravitational-wave sources, and in some cases merge to become Type Ia supernovae. So far, no DWD has been identified solely through relativistic Doppler beaming, even though the beaming amplitude directly relates to the radial velocity semi-amplitude. In this work, we initiate a comprehensive binary population synthesis using SeBa and incorporate the resulting binaries into a tripartite Galaxy model. Our proof-of-concept simulations demonstrate that the Vera C. Rubin Observatory Legacy Survey of Space and Time (LSST) can reliably recover relatively bright ($r \lesssim20~$mag) unequal-mass binaries in compact orbits with P $\approx$ 10--600 minutes with moderate to high inclinations. We find that LSST can detect at least 287 short-period DWDs, of which 47 are LISA-detectable gravitational wave sources. LSST lightcurves allow us to readily determine the period and fully characterize the orbit, in contrast with the challenges of orbit determination for DWDs in spectroscopic searches. The formation of unequal mass, short-period DWDs strongly depends on the assumptions regarding the mass-transfer phases during binary population synthesis, and the total number and characteristics of Doppler-beamed DWD systems observed in LSST will provide new tests of models of stellar binary evolution. Here, we lay the foundation for the comprehensive integration of synthetic Galactic binary population into realistic LSST survey simulations, thereby enabling quantitative forecasts of the number and characteristics of any binary sub-population during the LSST era.
\end{abstract}



\section{Introduction}

With nearly half of all stars in the Galaxy in binaries, stellar binary evolution plays a key role in shaping the visible universe. The interactions and mergers within stellar binary systems lead to some of the most exciting astrophysical transients \citep{pacz71,ropk23,marc24}. The mergers of two White Dwarfs (WDs) may dominate the production of Type Ia supernovae \citep[e.g.,][]{Iben_Tutukov1984,Webbink1984,maoz_observational_2014,ruiter_type_2025}, a cornerstone of modern cosmology \citep{riess_observational_1998,perlmutter_measurements_1999}. Compact binaries are sources of gravitational waves (GW) \citep{nelemans_gravitational_2001}, and double WD (DWD) binaries are expected to be the most dominant Galactic source of GW detectable by upcoming space based GW detector Laser Interferometer Space Antenna (LISA); \citealt{NelemansG+01_LISA,kupfer_lisa_2018,korol_prospects_2017,jin_detecting_2024}).

Discovery of DWDs in large-scale spectroscopic surveys such as Sloan Digital Sky Survey (SDSS) is inefficient due to the time intensive multi-epoch spectroscopic observations and low DWD binary fraction of $\approx$10\% \citep{badenes_merger_2012,maoz_binary_2017,adamane_pallathadka_double_2025-1}. So far, only around 400 close DWDs (optically unresolved) have been spectroscopically confirmed \citep{munday_dbl_2024}. However, only a fraction of these have measured periods, and an even smaller fraction with compact orbits with periods of few hours. The discovery of short-period DWDs in large-scale photometric surveys such as Zwicky Transient Facility \citep[ZTF;][]{BellmE+19_ZTF,DekanyR+20_ZTF} and Transiting Exoplanet Survey Satellite \citep[TESS;][]{RickerG+15_TESS} is possible only in edge-on systems that show eclipses due to the compact radii of WDs. Low-mass short-period WD binaries can be identified due to tidal deformations and the resulting lightcurve variations \citep{hermes_radius_2014}, while binaries detected through reflection effects require an exceedingly hot companion \citep[e.g.,][]{KupferT+19,KupferT+20_sdOB,SchaffenrothV+22} 


{

Most well-known close WD binaries have been identified through effects that are highly dependent on geometry or temperature \citep{Iben_Tutukov86a, YungelsonL+94,NelemansG+01_DWD,HanZ98,ToonenS+12_SeBa,ToonenS+17+WDPS}. Recently, \cite{HellstromL+25} show that eccentric, tight DWs form through dynamical interactions in dense stellar clusters, not in the field. While rare—about $10-15$ in the Milky Way—they can produce detectable GWs. These binaries offer a chance to independently measure their host distances clusters \citep[see,][]{WillemsB+07}.


An alternative pathway to discovering short-period DWDs is through the Doppler beaming effect, but it requires extremely precise photometry. Doppler beaming is a special-relativistic variation in the observed flux due to the time-varying velocity of the source relative to the observer \citep{zucker_beaming_2007}. Space missions such as \emph{CoRoT} and \emph{Kepler} have already demonstrated that Doppler beaming can be utilized to detect non-eclipsing compact binaries and to directly infer their radial velocity (RV) semi-amplitudes from photometric data \citep[e.g.,][]{Mazeh_Faigler10,EigmullerP+18,ZhengC+24}. \cite{shporer_ground-based_2010} reported the first ground-based measurement of Doppler beaming in a DWD with orbital period of 5.6 hr and beaming amplitude of only 3 mmag. Typically, Doppler beaming effect is only of the order of a few milli-mags or less, and WDs are faint. Consequently, to the best of our knowledge, no DWD binary has been discovered through Doppler beaming alone.

With the upcoming wide-field time-domain surveys with Vera Rubin Observatory’s Legacy Survey of Space and Time \citep[LSST;][]{LSST_Sci_Col09,LSST_Sci_Col17,IvezicZ+19_LSST, FantinN+20_LSSTWDs}, the single visit photometric error can be as low as 5 millimags, which is comparable to Doppler beaming amplitudes. These amplitudes are readily measurable when data from hundreds of visits are aggregated. LSST expands beyond targeted follow-up searches and covers the entire Southern sky, and thus we are finally at the precipice of detecting binaries through Doppler beaming alone. This sample will complement the several hundred DWDs that will be discovered in LSST using eclipses \citep{korol_prospects_2017,jin_detecting_2024}, will vastly increase the number of short-period DWDs, and contribute to the growing sample of LISA verification binaries -- DWDs that are detectable by LISA as individual sources of low-frequency GWs \citep{finc23}.


This study is similar to the works of \cite{korol_prospects_2017} and \cite{jin_detecting_2024}, which estimated the number of eclipsing binaries detectable by LSST. We integrate binary population synthesis with a realistic model of the Milky Way to predict the number and characteristics of DWDs that LSST can detect through Doppler beaming, employing the SeBa rapid binary evolution code \citep{NelemansG+01_DWD,ToonenS+12_SeBa,fantin_white_2020}. These systems are incorporated into a three-component Galactic model, encompassing the thin disk, the thick disk, and the bulge. We utilize a time-dependent star formation history and assign sky positions and distances accordingly. For these DWDs, we generate LSST light curves based on the baseline ten-year observing strategy and assess the efficiency of Doppler beaming in enabling the discovery of DWDs. Our primary objective is to evaluate whether LSST can identify approximately $\sim 10^2$ DWDs through beaming and develop a strategy to efficiently identify DWDs in LSST with minimal false-positives.

The paper is organized as follows: in Sec.~\ref{sec:DWD_MWPopSynth}, we describe the Galactic DWD population synthesis and spatial distribution; in Sec.~\ref{sec:LCSimul}, we present our LSST observing and period-recovery simulations; in Sec.~\ref{sec:results} and Sec.~\ref{sec:discussion}, we summarize our main results and discuss their implications for DWD demographics and future GW observations; and in Sec.~\ref{sec:conclusions} we provide our final conclusions. Orbital inclinations are defined relative to the plane of the sky. 
}

\section{Simulating Galactic DWD Population} \label{sec:DWD_MWPopSynth}

\subsection{Binary Population Synthesis}

To simulate the DWD formation in the Galaxy, we use the rapid binary population synthesis code SeBa \citep{NelemansG+01_DWD,toonen_effect_2013}. For the mass-transfer phases in binary evolution, we follow the $\gamma\alpha$ prescription and the parameter values assumed by \cite{toonen_effect_2013}. The primary star mass is drawn from the Kroupa mass function between 0.9 -- 10 M$_{\odot}$ with solar metallicity. The secondary to primary mass ratio is drawn uniformly between (0,1]. The semi-major axis is drawn from a power-law distribution with index -1 between 0 and $10^6~\mathrm{R_{\odot}}$, such that the semi-major axis distribution is flat in log scale. The eccentricity is drawn from a uniform distribution, which has been favored over the thermal distribution for close binaries \citep{raghavan_survey_2010,moe_mind_2017,hwang_eccentricity_2022,wu_eccentricities_2025}. We also repeated the analysis for a thermal distribution of eccentricity but found negligible difference in our final result, which is likely because the close binaries that form short-period DWDs quickly circularize, making the initial eccentricity largely inconsequential.

\subsection{Galactic Distribution of Stars}

\begin{figure*}
    \includegraphics[width=\linewidth]{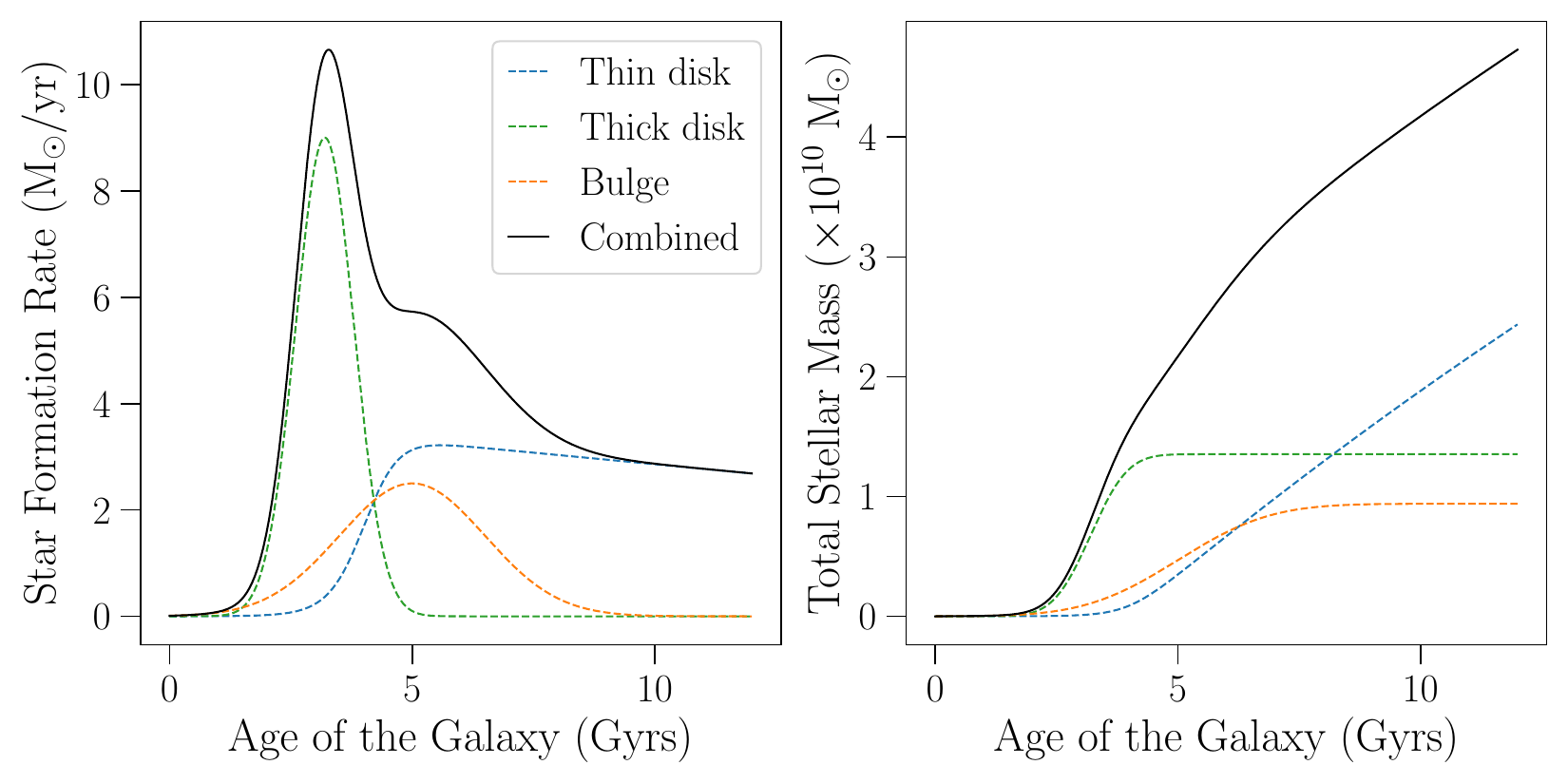}
    \caption{Left: Our adopted model for the SFR of the Milky Way (black line) composed of the thin disk (blue dashed line), thick disk (green dashed line), and the bulge (orange dashed line). Right: Cumulative stellar mass of these components as a function of time.}
    \label{fig:SFR}
\end{figure*}

We simulate the Galactic stellar population by dividing the Galaxy into the thin disk, the thick disk, and the bulge, as described in Appendix~\ref{sec:galactic_distribution_stars}. For the thin and thick disk, we model the star formation following \cite{fantin_canadafrance_2019} and approximate it using closed form functions, resulting in the star formation rate (SFR) as a function of time shown in Fig.~\ref{fig:SFR}. The current thin disk SFR is 2.6 M$_{\odot}$/yr, and the peak of the thick disk SFR at 9 M$_{\odot}$/yr and 0.6 Gyr width. For the bulge, we assume a Gaussian distribution centered at 4.5 Gyr (8.5 Gyr in lookback time) and 1.5 Gyr width. The normalization is chosen such that total bulge mass matches the results of \cite{licquia_improved_2015}. We obtain a total present-day stellar mass of $4.73\times10^{10}\mathrm{M_{\odot}}$. 

We split the age of the Galaxy into $dt=0.5~\text{Gyr}$ bins and calculate the mass of formed stars in each bin for each component. We assume half of all visible stars are in binaries, or an initial binary fraction of 0.5, consistent with past binary population synthesis studies \citep{toonen_supernova_2012,lamberts_predicting_2019,breivik_cosmic_2020}. \cite{moe_mind_2017} find the binary fraction to be less than 50\% and dependent on the primary mass, and a fraction of stars to be in triples and quadruples. However, we only simulate binaries following the binary population synthesis simulations in the literature and, thus, set a constant binary fraction of 0.5 for simplicity. Under this assumption, the total mass of stars in primaries is the same as the total mass of single stars in the galaxy. Assuming a uniform mass-ratio for the secondary, over a large sample (which is always satisfied at galactic scales) the average mass ratio is 0.5. Thus, for the entire Galactic stellar population 40\% of the mass is in single stars, 40\% of the mass is in the primary stars of the binaries, and finally 20\% of the mass is in the secondaries. The 40\% of the mass in the primaries is distributed between [0.1,100] M$_{\odot}$ based on Kroupa mass function. By integrating the Kroupa mass function (or equivalently, by drawing a large mock sample), we find that nearly half of the total mass is contained in our simulation range between [0.9,10] M$_{\odot}$ -- i.e., in the types of stars that can lead to the formation of WDs in binaries over cosmic time. Thus, we expect 20\% of the total stellar mass is assigned to be in such primaries in our simulation. Further, the mean mass of the stars sampled between [0.9,10] M$_{\odot}$ is 2.11 M$_{\odot}$. If $d{M_{T}(t)}$ is the total stellar mass formed in a time bin $[t,t+dt]$, then we expect nearly $d\mathrm{N(t)}$ = (0.2$d{M_{T}(t)}$)/2.11 number of such binaries. 
For each star, we assign a formation time uniformly within the time bin $[t,t+dt]$ with a resolution of 10 million years. The orbital inclination for each binary is drawn from a  sinusoidal distribution, consistent with randomly oriented orbits.

Simulating the entire Galactic population, which amounts to nearly 4 billion binaries in the range of parameters that is of interest to us, is computationally expensive. The stellar populations of a given age in different regions of the Galaxy are identical in our model. Thus, we simulate 50 million binaries in SeBa, with stellar evolution time inferred based on the star formation distribution described above. We then assign coordinates and formation time to 1/4th of the total Galactic population, or nearly 1 billion binaries and resample the 50 million simulated binaries to a much bigger sample of 1 billion. While the same DWD will get assigned multiple times in the final sample, these copies will be distributed across the Galaxy and still give us a well-sampled population. Finally, any number that is inferred from this population is multiplied by 4 to get the total realistic number estimates.

\subsection{Synthetic photometry}

We use synthetic WD colors to simulate the observed photometry \citep{kowalski_found_2006,tremblay_improved_2011,blouin_new_2018,bedard_spectral_2020}\footnote{\href{https://www.astro.umontreal.ca/~bergeron/CoolingModels/}{\url{https://www.astro.umontreal.ca/~bergeron/CoolingModels/}}}. LSST filters are well approximated by SDSS and Pan-STARRS optical filters, and we use SDSS \textit{u-}band absolute magnitude to represent the LSST \textit{u} band, and Pan-STARRS \textit{grizy} absolute magnitudes for the LSST \textit{grizy} bands. The absolute magnitude of the binary is computed from the sum of fluxes from both WDs. Taking into the account the dust extinction $A_b$ (calculated as described in Appendix \ref{sec:dust_extinction}) for a DWD at distance $d_{\text{kpc}}$, the observed magnitude in a band $b$ can be written as:

\begin{equation}
    m_{\text{b,obs}} = m_{\text{b,abs}} + 5\log d_{\text{kpc}} + 10 + A_b.
\end{equation}

For this work, we limit our analysis to targets brighter than 20th magnitude. Such WDs are in the \textit{Gaia} footprint with good distance estimates, and we use the nearly complete sample of WDs for $G<20$ mag by \cite{gentile_fusillo_catalogue_2021} to estimate the false-positive rate. Our work can be extended once the WD population is better characterized at fainter magnitudes. We also discard systems brighter than the 17th magnitude as they are near the saturation limit of the LSST detector.

\section{Lightcurve Simulation and Period Recovery} \label{sec:LCSimul}
\subsection{Simulating LSST lightcurves}

To simulate the LSST cadence, we use the 10-year simulated LSST observations with the baseline strategy \texttt{baseline\_v5.0.0\_10yrs} \footnote{\url{https://s3df.slac.stanford.edu/data/rubin/sim-data/sims_featureScheduler_runs5.0/baseline/}}. For each DWD, we select all fields that are within the LSST FOV (1.75$\degree$ radius). A source in LSST footprint typically gets about 700 observations with a minimum time gap of 30 minutes between exposures. 

The Doppler beaming amplitude at band-pass wavelength $\lambda_{\text{b}}$ for a binary with orbital period $P$, masses $m_1$ and $m_2$, surface temperatures $T_{\mathrm{eff,1}}$ and $T_{\mathrm{eff,2}}$, observable RV semi-amplitudes $K_1$ and $K_2$, blackbody fluxes $F_{\lambda_b,1}$ and $F_{\lambda_b,2}$, and observed magnitude $m_{\text{b,obs}}$, is given by \citep{zucker_beaming_2007}:
\begin{equation}
    A(\lambda_{\text{b}},m_{\text{b,obs}}) = \frac{1}{c} \frac{K_1\alpha_1^{\prime}(\lambda_{\text{b}})F_{\lambda_{\text{b}},1}-K_2\alpha_2^{\prime}(\lambda_{\text{b}})F_{\lambda_{\text{b}},2}}{F_{\lambda_{\text{b}},1}+F_{\lambda_{\text{b}},2}}.
    \label{equation:zuker_2007_doppler}
\end{equation}
Here, $\alpha_{i}^\prime(\lambda) = y_{i}e^{y_{i}}/\left(e^{y_{i}}-1\right)$ and $y_{i} = hc/\left(\lambda_{\text{b}} k_B T_{\mathrm{eff,{i}}}\right)$, with $i = 1,2$.

Assuming circular orbits, as expected for systems that have experienced common-envelope evolution and subsequent GW circularisation, and including the observational error $\delta m_{\text{b,obs}}$, the binary model above results in a time-varying flux given by
\begin{equation}
    \frac{\Delta f(t)}{f(t)} \approx \Delta m(t) = A(\lambda,m_{\text{b,obs}}) \sin(2\pi\frac{t}{P} + \phi) + \delta m_{\text{b,obs}}.
\end{equation}
The standard deviation of the normally distributed error of LSST photometry can be expressed as \citep{IvezicZ+19_LSST}:
\begin{equation}
    \sigma_b = \sqrt{\sigma_{sys}^2 + \sigma_{rand}^2}\ ;\ \ \ \ \sigma_{rand}^2 = (0.04-\gamma)x + \gamma x^2 ,
\end{equation}
where the systematic error is $\sigma_{sys} = 0.005~\text{mag}$ and the random error is $\sigma_{rand}^2$. Here, $x=10^{0.4(m_{\text{b,obs}}-m_5)}$, $m_5$ is the $5\sigma$ depth obtained from the baseline simulation and $\gamma$ depends on sky brightness, readout noise, etc., and is taken from Table 2 of \cite{IvezicZ+19_LSST}.

\begin{figure}
    \centering
    \includegraphics[width=\linewidth]{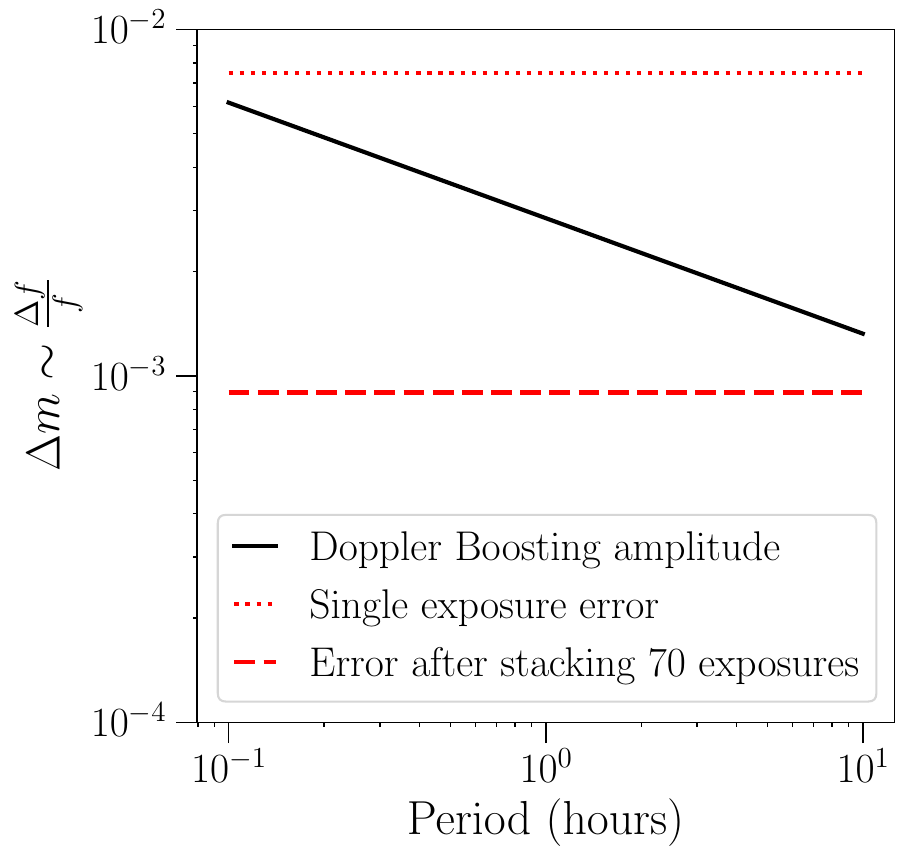}
    \caption{Doppler beaming amplitude is shown for a binary with $\mathrm{(M_1,T_{eff,1})}$ $\mathrm{=(0.3~M_{\odot},10000~K)}$, $\mathrm{(M_2,T_{eff,2})}$ $\mathrm{= (0.6~M_{\odot},7500~K)}$, inclination of 60 degrees, and source magnitude of $20~\mathrm{mags}$ in $r$ band. The dotted line corresponds to a single exposure error, and the dashed line is the estimated error after stacking 70 exposures or approximately 10\% of the total LSST exposures for a typical target. 
    }
    \label{fig:dopplerBeaming}
\end{figure}

In Fig.~\ref{fig:dopplerBeaming} we show the variation resulting from Doppler beaming for different orbital periods for a prototypical system. The single-exposure photometric error exceeds the Doppler beaming amplitude and is insufficient to detect photometric variation. However, over the lifespan of LSST such a system will be observed up to approximately 700 times covering several thousand orbits. Performing a periodogram search has a similar to effect to stacking the observations at the same phase, and we can use this technique to check whether the system will be visible to LSST as a binary. Stacking 70 exposures, or 10\% of all observations, reduces the photometric error as shown in Fig.~\ref{fig:dopplerBeaming}. Thus, in the phase-folded lightcurve the Doppler beaming amplitude can be greater than the photometric error, and, under favorable conditions, LSST data can be used to classify the system as a binary.

\subsection{Period Recovery:}
We use the accelerated Lomb-Scargle periodogram implemented by \texttt{NIFTy} in \texttt{astropy} to find the best-fit period \citep{lomb_least-squares_1976,scargle_studies_1982,garrison_nifty-ls_2024}. We perform the periodogram search for periods between 10 minutes and 50 hours, a typical range of periods expected for photometric variation in WDs. The Doppler beaming amplitude can vary for different bands, and some bands, such as the $r$-band, receive much more observations than others. We combine data from all the photometric bands into a single lightcurve. These differences in amplitudes and the effect of combining them can be seen as tracks with differing amplitudes in the lightcurves shown in Fig.~\ref{fig:period_recovery}. In principle, performing a multi-band Lomb-Scargle periodogram will produce better results (see, e.g, \citealt{vand15}). However, we find this computationally impractical and suggest it only for improving systems that are first identified as varying. 

We compute the false-alarm probability in \texttt{astropy} which is the probability that a noisy dataset with the same cadence produces a periodic power as strong as the best-fit period \citep{baluev_assessing_2008}. We classify systems with false alarm probability less than $10^{-5}$ as binary candidates. \textit{Gaia} has observed around 200,000 WDs brighter than the 20th magnitude in optical band, which is a nearly complete sample \citep{gentile_fusillo_catalogue_2021}. Only about half of these are in the southern hemisphere and fall in the LSST footprint. A false alarm probability of 0.001\% implies that we expect around one WD to be incorrectly classified as a binary. We find that increasing this cutoff does not significantly increase the number of DWDs detected, while the false positives increase rapidly. Considering systems beyond 20th magnitude can also increase the number of DWDs detected, but at the cost of increased number of false-positives. A better characterization of LSST noise behavior in real observations, combined with simulations of the single WD population, will allow us to make a less conservative estimate on the number of DWDs detected in the future. 

\section{Results}  \label{sec:results}
\begin{figure*}
    \centering
    \includegraphics[width=0.49\linewidth]{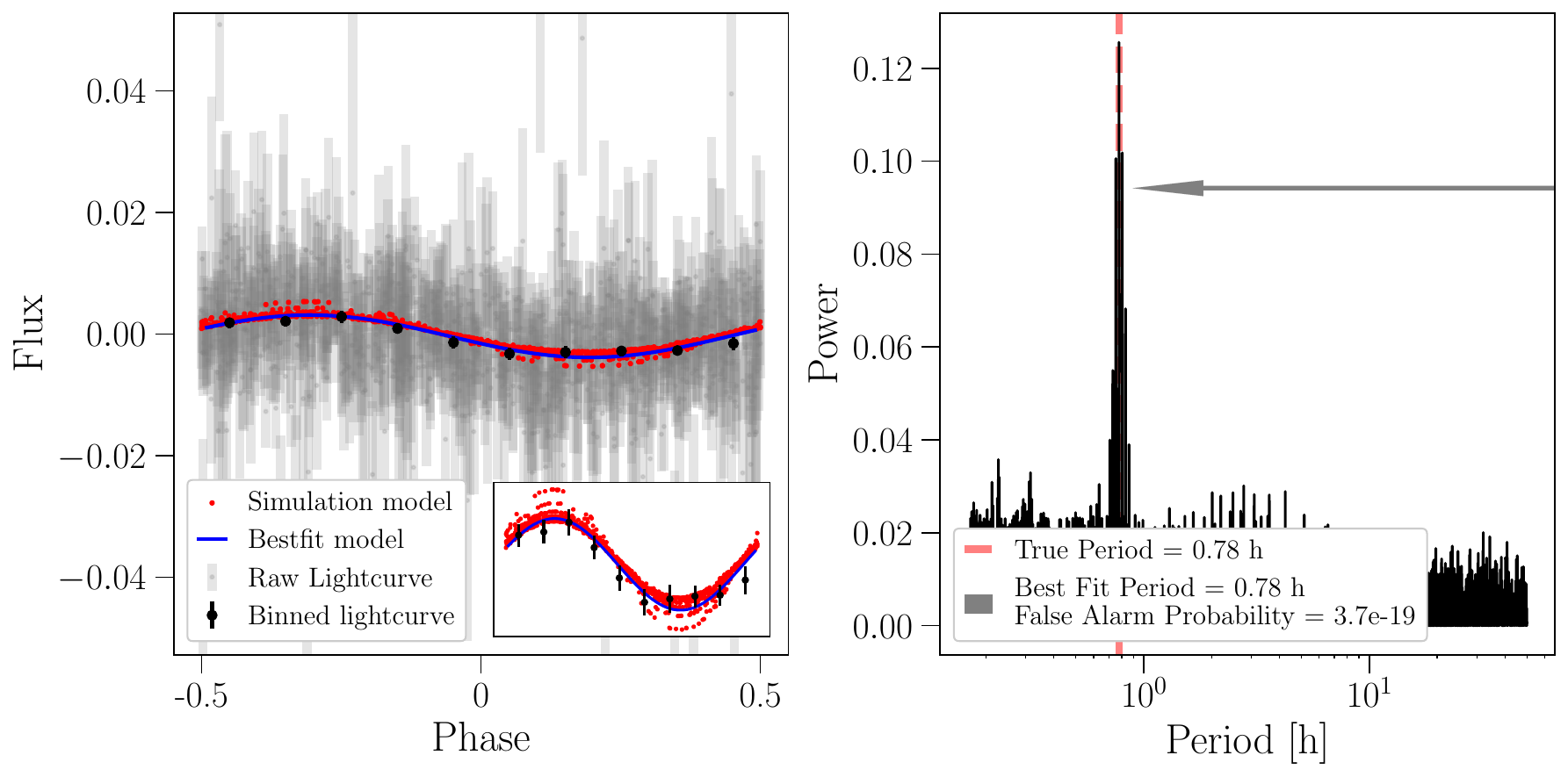}
    \includegraphics[width=0.49\linewidth]{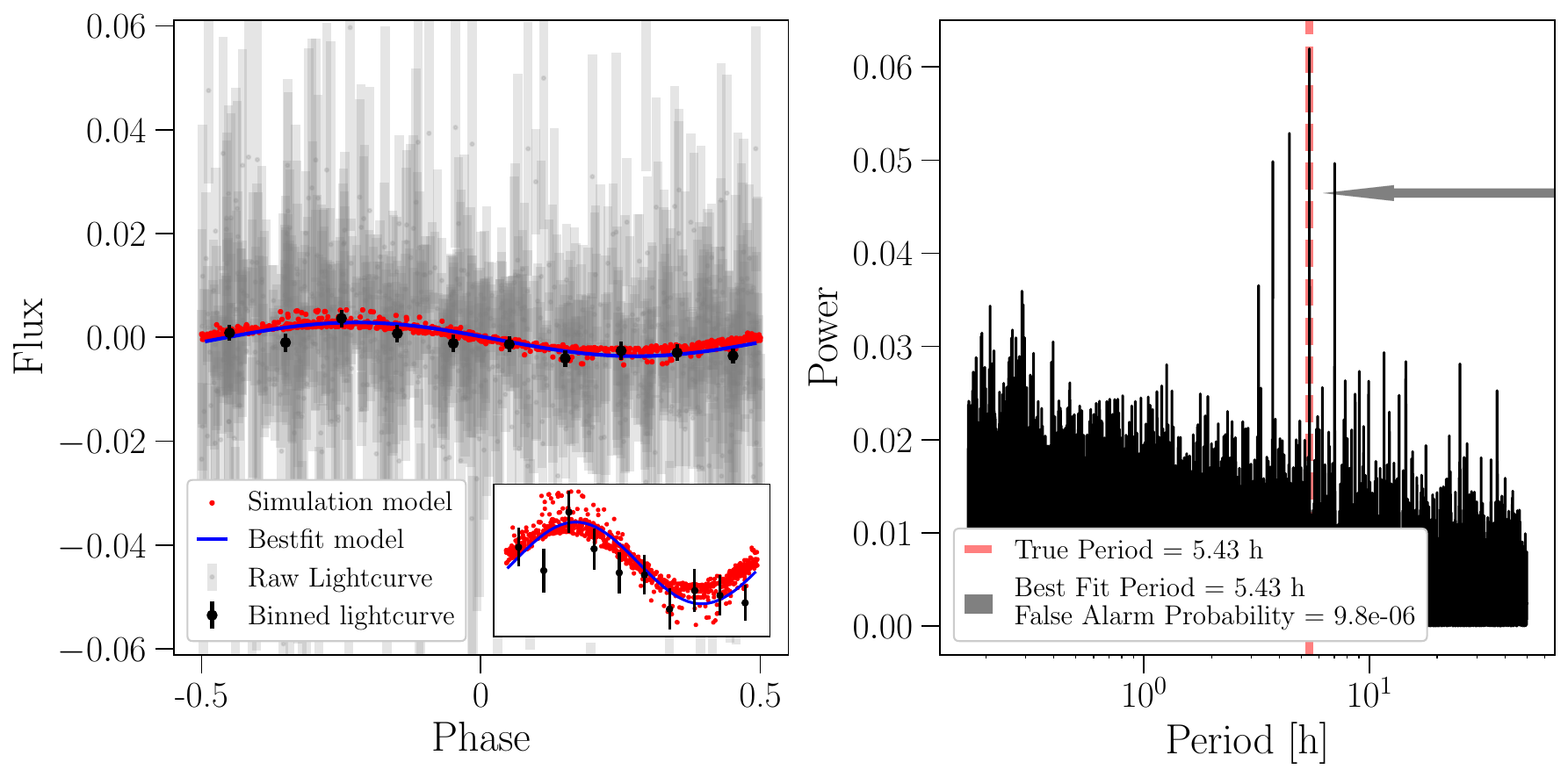}
    \includegraphics[width=0.49\linewidth]{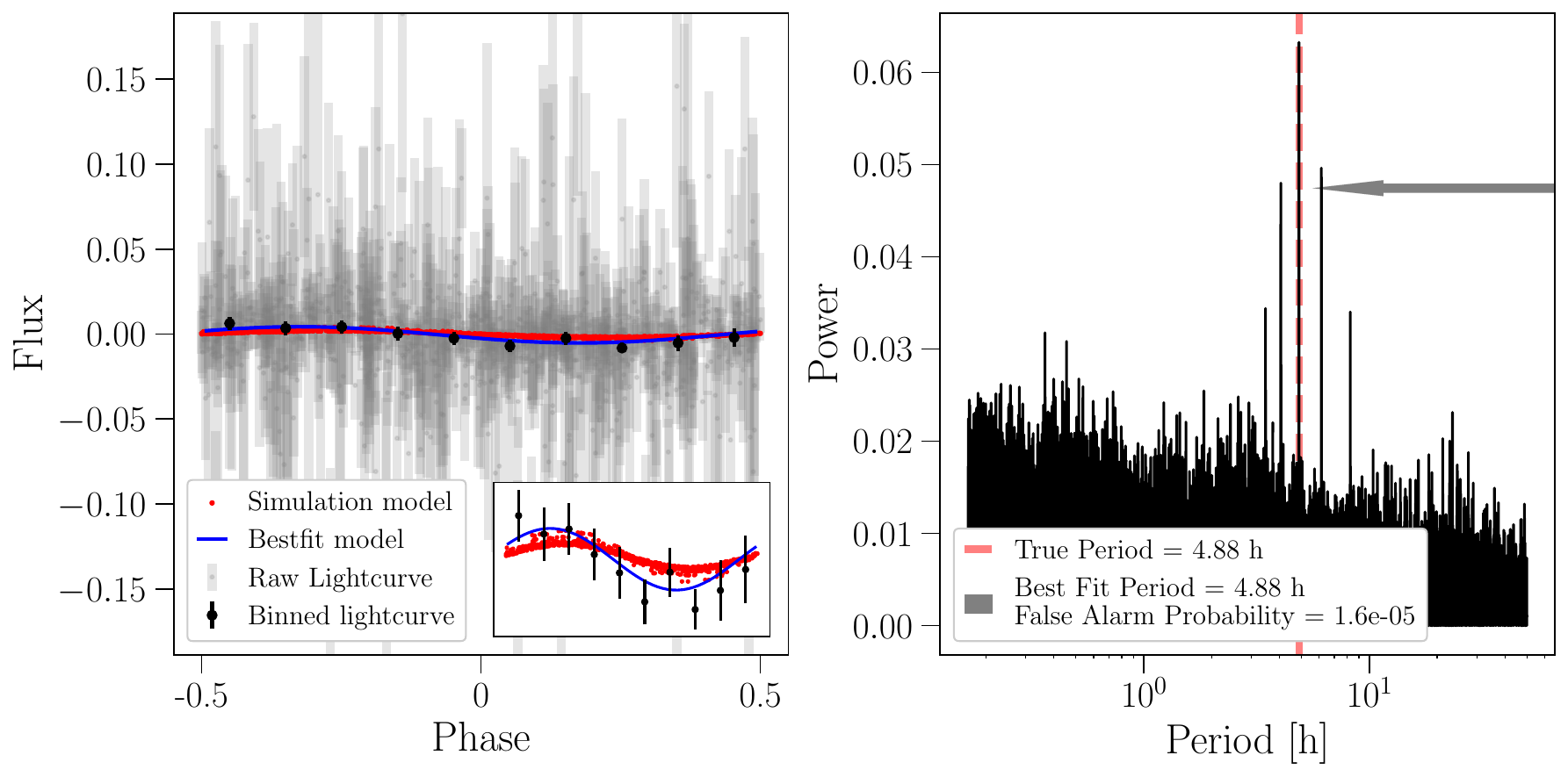}
    \includegraphics[width=0.49\linewidth]{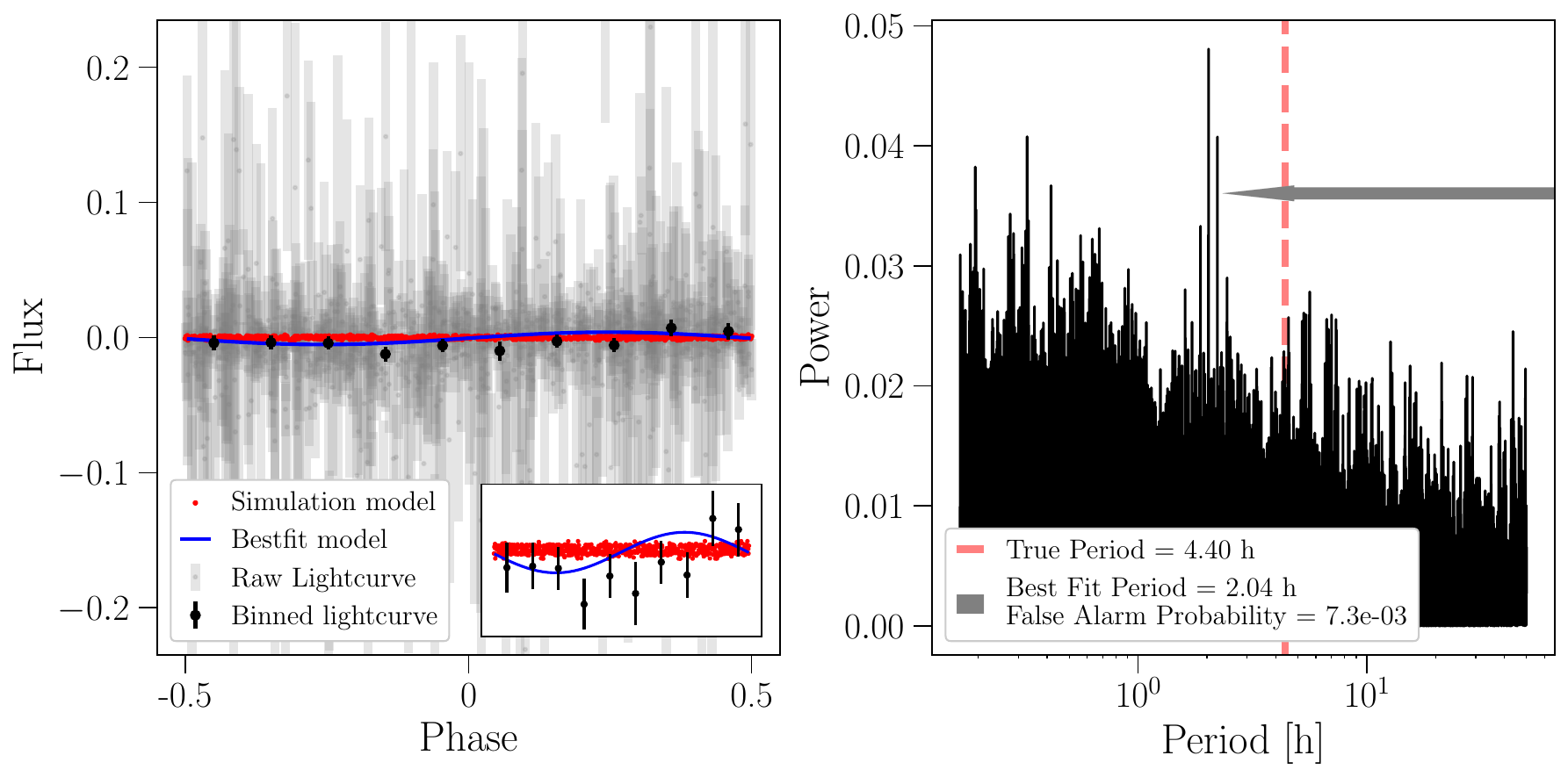}
    \caption{Examples of period recovery for different values of false alarm probability. The systems in the top row would be classified as binary candidates, while those in the bottom row would be discarded based on our chosen cutoff. We have combined lightcurves from all available bands to reduce computational cost of periodogram search, even if the beaming amplitude varies for different bands, which can be seen in the inset plots as tracks with differing amplitudes.}
    \label{fig:period_recovery}
\end{figure*}

We repeat the lightcurve simulation and period recovery ten times for different random seeds to ensure that the results are largely independent of the choice of the initial seed. Taking the average of ten runs, we find that LSST can detect at least 287 DWD binary candidates using Doppler beaming alone, of which at most a couple will be false-positives. We find that for a vast majority of these we can recover the true period, although there can be secondary peaks confounding the period determination. In Fig. \ref{fig:period_recovery} we show four different examples of period recovery. The systems classified as binaries show a clear peak in the power spectrum, and the best-fit model closely matches the simulated model. As the false alarm probability increases, the amplitude of variation decreases and the binned lightcurve becomes noisier.

To estimate the number of sources detectable by LISA we calculate the characteristic strain following \cite{kupfer_lisa_2018} and compare it to the 4-year LISA sensitivity curve given by \cite{robson_construction_2019}. We find 47 systems that are detectable by LISA. With further followup, such systems will enlarge the number of LISA verification binaries. The result is summarized in Table~\ref{tab:result}.

\begin{table}
    \centering
    \begin{tabular}{c|c}
        Prediction type & Number \\
        \hline
        Total number of DWDs in the Galaxy & $7.2\times10^{8}$\\
        Number of DWDs identified in LSST & $287$\\
        Number of LISA detectable DWDs in LSST & $47$\\ 
    \end{tabular}
    \caption{Summary of the simulation and the number DWDs detected through Doppler beaming in this work.}
    \label{tab:result}
\end{table}

\section{Dicussion}
\label{sec:discussion}
\begin{figure*}
    \centering
    \includegraphics[width=0.9\linewidth]{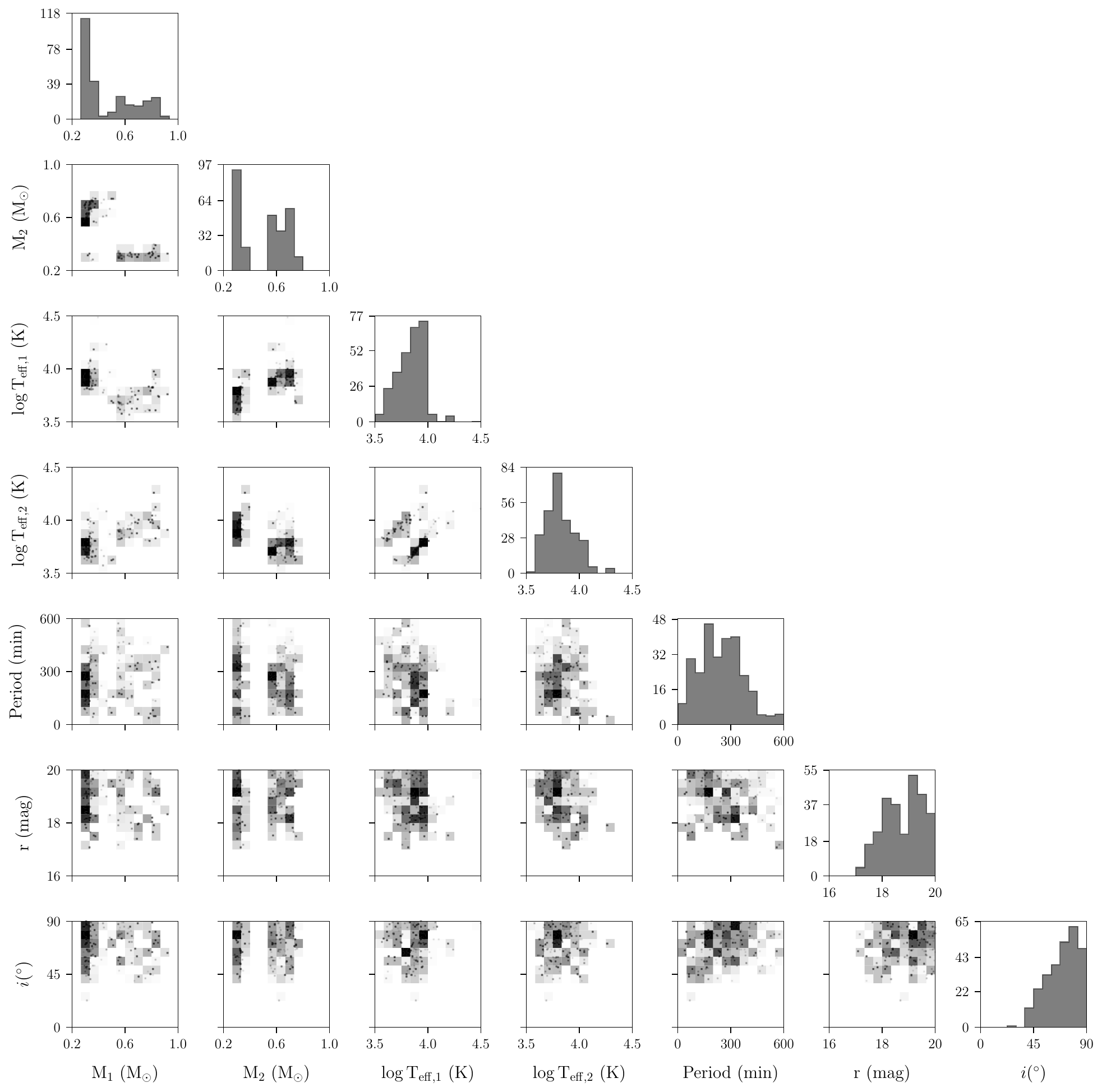}
    \caption{The distribution of parameters of recovered DWDs is shown. We show the 2D histograms in the background and the actual data points as a scatter plot on top.}
    \label{fig:corner_stellarParams}
\end{figure*}

Fig.~\ref{fig:corner_stellarParams} shows the inter-dependencies between various parameters of the detected DWDs. We find that most DWDs have a He-core WD companion, consistent with observations \citep{brown_elm_2020,kosakowski_elm_2023,adamane_pallathadka_double_2025-1}. The distribution of masses and temperatures makes it clear that Doppler beaming is strongest in systems where one of the two stars dominates the total luminosity, such that the Doppler beaming of one does not offset the Doppler beaming of the other. We find that shorter orbital periods are preferred but with a long tail, and with longer orbital periods requiring larger inclinations and brighter systems, as expected. 

\begin{figure*}
    \centering
    \includegraphics[width=0.32\linewidth]{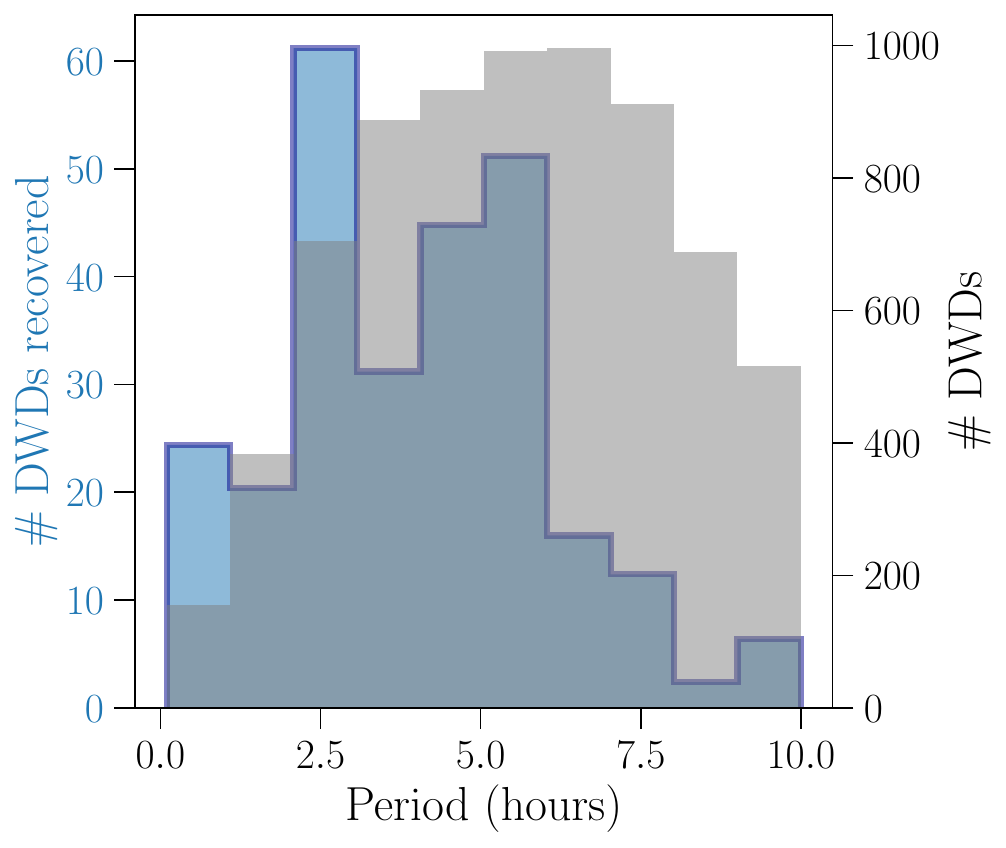}
    \includegraphics[width=0.32\linewidth]{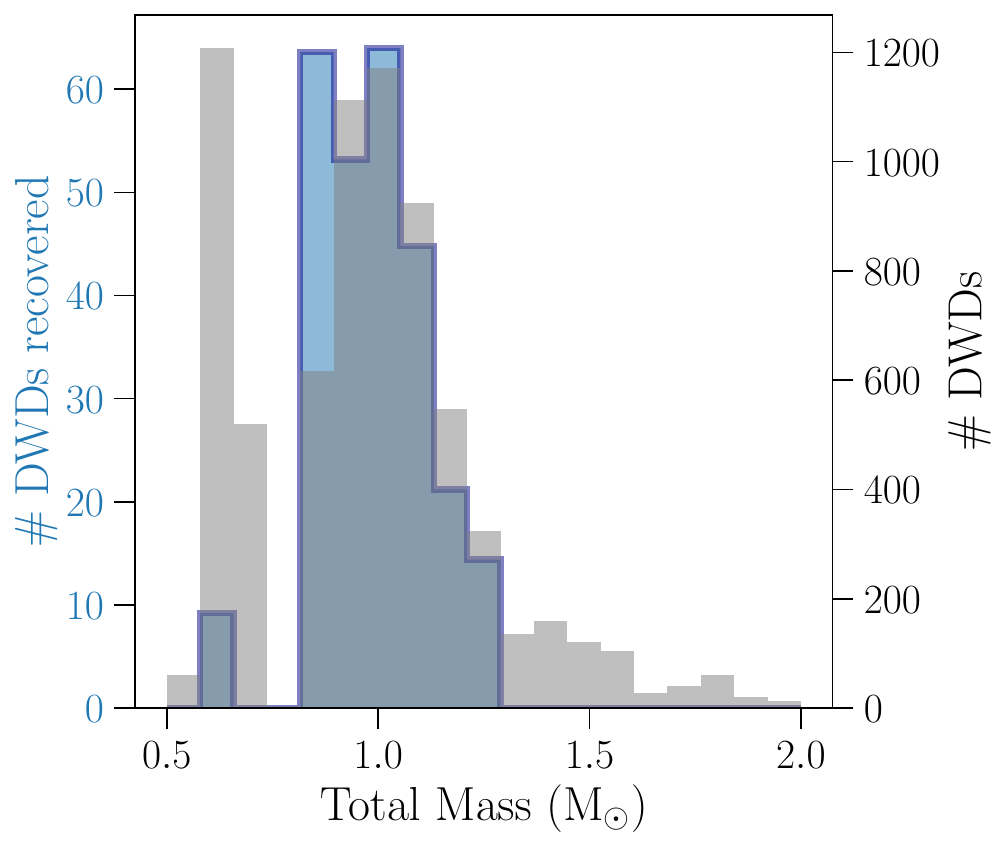}
    \includegraphics[width=0.32\linewidth]{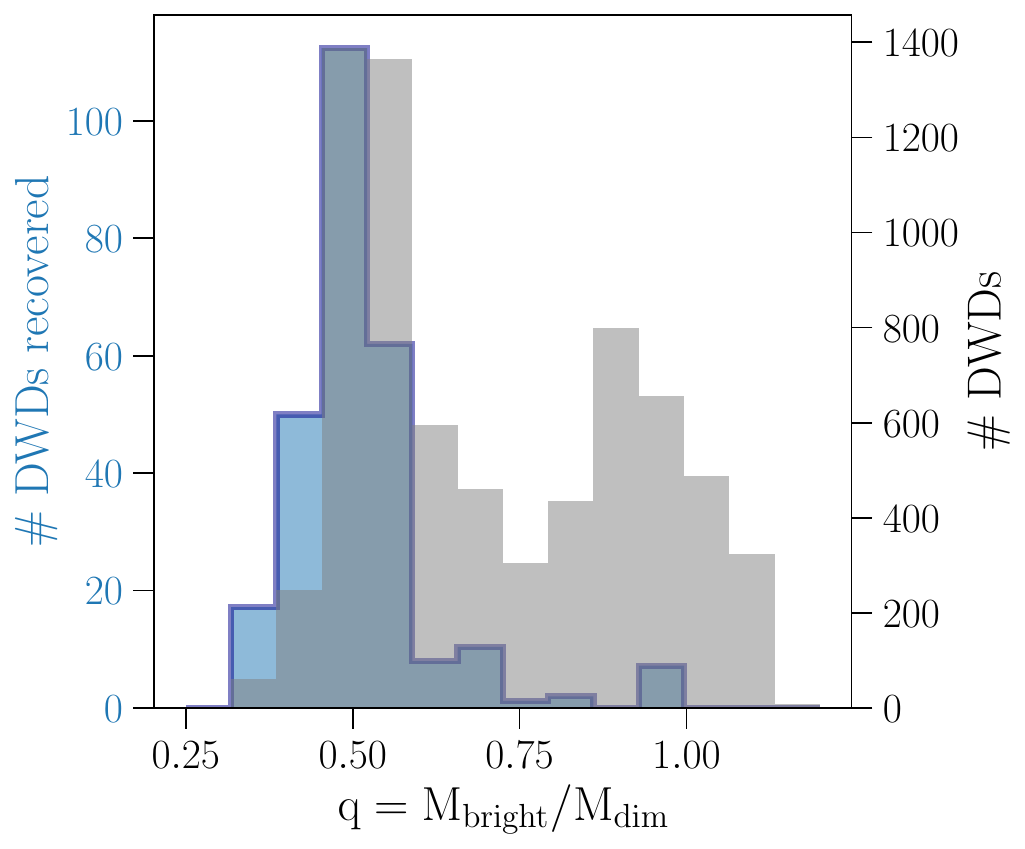}
    \caption{The distribution of periods (left), total mass (middle), and the mass ratio (right) of the recovered DWDs (blue) as compared to the rest of the DWDs with $r<20~$mag and period within 10 hours (grey). The figure illustrates the biases in the DWD sample selected through the Doppler beaming effect. We find that unequal mass binaries with period between 2--5 hours and total mass of 1 M$_{\odot}$ are preferentially recovered.}
    \label{fig:comparison}
\end{figure*}

To illustrate the multi-dimensional selection function that will influence the LSST detections, in Fig.~\ref{fig:comparison} we compare the period, total mass, and the mass ratio of detected DWDs to the rest of the DWDs with $r<20$ mag and orbital period less than 10 hours. The period distribution shows that short-period binaries with orbital periods shorter than 5 hours are preferentially selected -- the beaming amplitude is proportional to RV semi-amplitude which decreases at longer orbital periods. From the mass ratio distribution, it is clear that Doppler beaming detections are biased to unequal-mass binaries. This is primarily because the WDs in equal mass binaries will have similar RV semi-amplitudes, and due to their similar sizes, they are more likely to have comparable luminosities. Thus, the Doppler beaming of one offsets the other (see, Equation~\ref{equation:zuker_2007_doppler}). Also, at the same orbital period and same total system mass, the lower mass WD in unequal mass binaries has a larger RV semi-amplitude than either WDs in equal mass binaries, leading to a larger beaming amplitude. This preference for unequal mass binaries leads to a narrow peak in total mass around 1 M$_{\odot}$, since extremely low mass and extremely high mass systems are more likely to be produced by equal mass binaries. 

We can make use of these selection preferences to explore the mass-transfer phases of compact binary evolution. The mass-ratio distribution of DWDs strongly depends on assumptions regarding the mass-transfer phases during binary evolution. Some binary population models or the choice of parameters such as the $\alpha_{\mathrm{CE}}$ parameter of the common envelope phase can result in a dearth of unequal-mass binaries \citep{li_influence_2023}. This leads to nearly no binaries showing Doppler beaming signals. Thus, by comparing the total number of binaries detected through Doppler beaming in LSST and their characteristics with the binary population synthesis results, we can constrain the binary evolution models.

\begin{figure}
    \centering
    \includegraphics[width=0.65\linewidth]{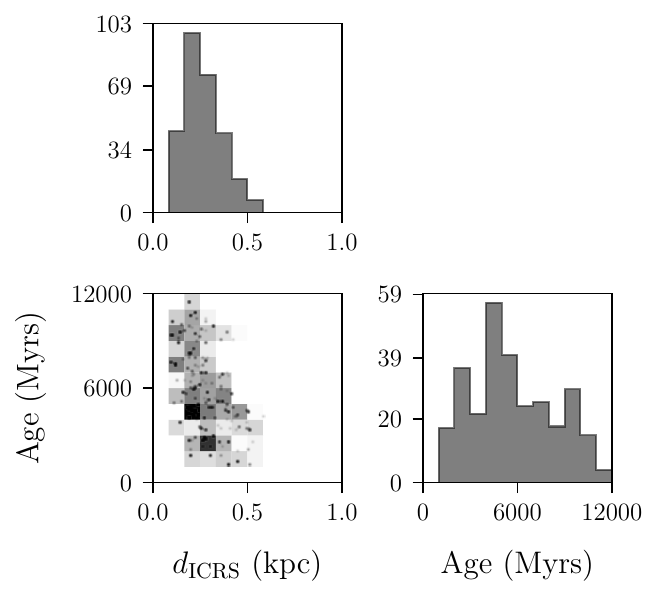}
    \caption{The distribution of age and distance of the detected binaries is shown. We find that younger binaries that are hotter than the rest are detected to farther away distances compared to rest of the sample.}
    \label{fig:corner_dzage}
\end{figure}

\begin{figure}
    \centering
    \includegraphics[width=0.95\linewidth]{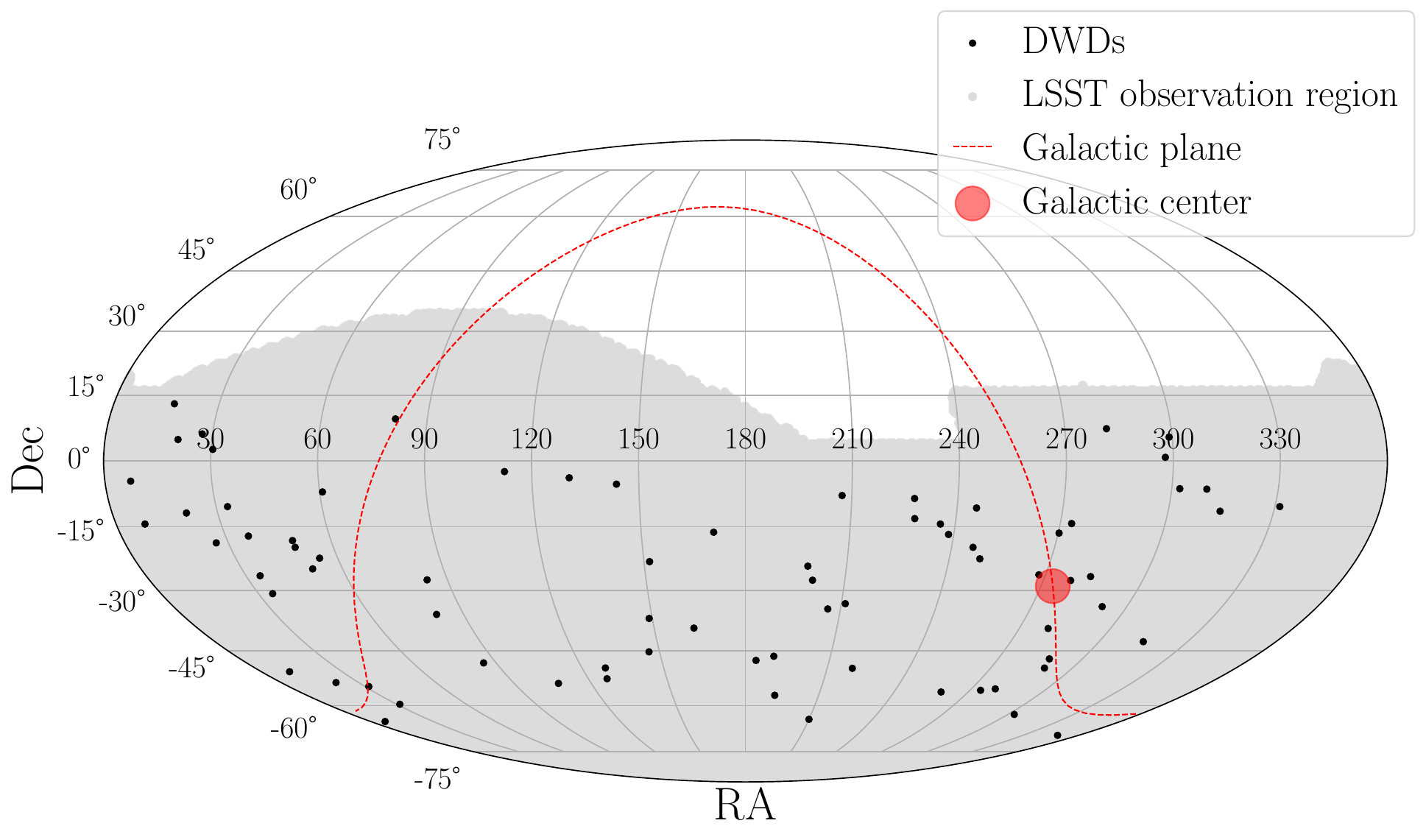}
    \caption{The distribution of detected DWDs in the Galaxy from one iteration of the period recovery simulation.}
    \label{fig:galactic_distribution}
\end{figure}

Fig.~\ref{fig:corner_dzage} shows that only systems within $\approx$ 500 pc are recovered, which is a consequence of our magnitude cutoff ($r<20$ mag). We find that most detected DWDs have ages around 6 Gyrs, indicating a thin disk origin, and a broad tail indicating a thick disk origin. Thick disk stars would originate with metallicity lower than the assumed solar value and consequently lead to WD masses greater than those with solar metallicity progenitors \citep{romero_age-metallicity_2015}. Thus, our work likely underestimates the number of massive WDs in binaries. Given the typical distances, we do not expect any stars to be located in the central bulge. The age-distance distribution shows a feeble tail, with stars further away being younger -- a consequence of them being hotter and brighter. These DWDs are distributed throughout the Galaxy, although there is a slight preference towards the Galactic center as seen in Fig.~\ref{fig:galactic_distribution}. 

Detecting DWDs through Doppler beaming can potentially be contaminated by several types of interesting interlopers. Pulsating WDs can produce lightcurve variations that may be confused with the periodic variability of Doppler-beamed binaries. However, pulsating WDs are restricted to a narrow range of stellar temperatures and to periods within a couple of hours \citep{van_grootel_instability_2012}. In addition to pulsating WDs, reflection binaries and ellipsoidal variables can produce similar sinusoidally varying lightcurves \citep{hermes_radius_2014}. Both these effects originate due to binary nature of the system and are equally interesting as Doppler beaming to detect short period DWDs. The color information provided by LSST and the spectroscopic followups will be helpful in distinguishing between these effects. 

\section{Conclusions} \label{sec:conclusions}
In this work, we have developed a forward-modeling framework to predict the binary population of DWDs that the LSST can detect through Doppler beaming. We start with a large SeBa binary population synthesis run, which provides the intrinsic properties and evolutionary histories of millions of binaries over 13 Gyrs. These systems are then embedded in a three-component Milky Way model (thin disk, thick disk, and bulge) with a time-dependent star formation history inspired by recent measurements of star formation rate. Together with a realistic spatial distribution and dust extinction model, this approach yields a synthetic present-day Galactic DWD population with distances and physical parameters suitable for comparison with observations, LSST positions, and magnitudes. With such a sample, we simulate their lightcurves including Doppler beaming and attempt to recover the binaries. We find that this is a promising way to expand the catalog of close DWD binaries and find that we can expect to observe at least 287 short-period DWDs with a median period of 4 hours, of which 47 will be LISA verification binaries. Once such systems are identified, inexpensive follow-up campaign of couple of exposures per target will be sufficient to determine the source of photometric variation and to classify the system as a binary, with a well-measured orbital period using the lightcurve. Observing and characterizing this population of binaries will help us answer some of the long-standing problems in stellar binary evolution.

\begin{acknowledgments}
GAP and NLZ acknowledge support by the JHU President’s Frontier Award to NLZ. YZ acknowledges support from MAOF grant 12641898 and visitor support from the Observatories of the Carnegie Institution for Science, Pasadena CA, where part of this work was completed. This work was carried out at the Advanced Research Computing at Hopkins (ARCH) core facility (rockfish.jhu.edu), which is supported by the National Science Foundation (NSF) grant number OAC 1920103.
\end{acknowledgments}

\begin{contribution}

GAP was responsible for writing and submitting the manuscript. All other authors contributed equally and edited the manuscript.


\end{contribution}

%


\section*{Data Availability}
All codes and final results of the DWD population, including the code used to produce the figures, are made public and available at the GitHub repository \href{https://github.com/ap-gautham/DopplerBeamingDWDs_LSST/tree/main}{DopplerBeamingDWDs\_LSST}.\\

\software{
\texttt{SeBa} \citep{ToonenS+12_SeBa}, \texttt{AstroColour} \citep{Lane_2025_colour}, \texttt{astropy} \citep{Astropy_2013, Astropy_2018, Astropy_2022}, \texttt{matplotlib} \citep{Hunter_2007}, \texttt{numpy} \citep{Harris_2020}, \texttt{pandas} \citep{McKinney_2010}, \texttt{scipy} \citep{Virtanen_2020}, \texttt{lightkurve} \citep{lightkurve_collaboration_lightkurve_2018},
\texttt{NIFTy} \citep{garrison_nifty-ls_2024},
\texttt{dustmaps} \citep{green_dustmaps_2018}.        }


\appendix

\section{Structure of the Galaxy}
\label{sec:galactic_distribution_stars}

We consider a realistic Galactic model by splitting the Galaxy into three components: thin disk, thick disk, and bulge \citep{frankel_measuring_2018,wagg_gravitational_2022}. Each component is modeled using cylindrical coordinates $(r,\phi,z)$.

\begin{enumerate}
    \item Dependence on $r$: Assuming exponentially decaying surface brightness profile of galaxies, the probability density distribution of stars is given by:
    \begin{equation}
    p(r) = \frac{r}{r_s^2}\exp{\bigg(-\frac{r}{r_{s}}\bigg)},
    \end{equation}
    where $r_s$ is the scale-length and varies for each Galactic components. This is a gamma distribution with shape parameter 2 and scale parameter $r_{s}$. We make use of \texttt{scipy.stats.gamma} to draw samples from this distribution.
    \item Dependence on $z$: The exponentially decaying surface brightness profile assumption results in a probability density distribution of $z$-coordinate of stars give by:
    \begin{equation}
    p(z) = \frac{1}{z_s}\exp{\bigg(-\frac{|z|}{z_{s}}\bigg)},
    \end{equation}
    where $z_s$ is the scale-length. This is a gamma distribution with shape parameter 1 and scale parameter $z_{s}$. We again make use of \texttt{scipy.stats.gamma} to draw samples from this distribution.
    \item Dependence on $\phi$: The angular component is uniformly assigned between [0,2$\pi$]. 
\end{enumerate}

The full 3D probability distribution is the product of the three probability densities. The different scale-lengths are summarized in Table \ref{tab:scale_length}. The coordinates $(r,\phi,z)$ are converted to Galactocentric coordinates $(x,y,z) \equiv (r\cos\phi,r\sin\phi,z)$, which are then converted to (RA,dec,$d$) using \texttt{astropy}. For the Galactocentric coordinates, we use the \texttt{astropy} default definition with the Earth situated along negative $x$-axis i.e $\phi=$180 degrees.

\begin{table*}
    \centering
    \begin{tabular}{c|c|c|c}
        Component & $r_s$ & $z_s$ & Reference\\
        \hline
        Thin Disk ($H/R \simeq 0.2$) & $\mathrm{3~kpc\times(1-0.3\frac{(13-t/Gyr)} {8})}$ & 0.3 kpc & \cite{bland-hawthorn_galaxy_2016,frankel_measuring_2018}\\
        Thick Disk ($H/R \geq 0.3$)& 2 kpc & 0.9 kpc & \cite{bland-hawthorn_galaxy_2016}\\
        Bulge ($H/R \geq 0.3$) & 0.5 kpc & 0.2 kpc &\cite{bland-hawthorn_galaxy_2016}
    \end{tabular}
    \caption{Distribution of parameters of the Galactic components. Here, $H/R \sim c_s /\left(r\Omega(r)\right)$.}
    \label{tab:scale_length}
\end{table*}

\section{Dust extinction}
\label{sec:dust_extinction}
Following the same prescription as \cite{sandage_redshift-distance_1972} and \cite{nelemans_short-period_2004}, we assume a uniform sheet of dust along the Galactic plane with density in the $z$-direction proportional to the stellar density of the thin disk, with scale height $z_s = 0.3~\text{kpc}$. This approximation works well when the observed stars are at a distance smaller than the scale-lengths in Table~\ref{tab:scale_length}, which is true for WDs  with $r \lesssim 20~\text{mag}$. Under this assumption, extinction can be written as the ratio of the total column density of dust towards a  star with Galactic coordinates $(l,b)$ at distance $d$ and a star along the same direction at infinite distance: 
\begin{align}
\begin{split}
    \frac{A_V(l,b,d)}{A_V(l,b,\infty)} &= \frac{\int_0^{d}\exp{\big(-\frac{L\sin |b|}{z_s}\big)\text{d}L}}{\int_0^{\infty}\exp{\big(-\frac{L\sin |b|}{z_s}\big)\text{d}L}} \\
    \implies A_V(l,b,d) &= A_V(l,b,\infty) \bigg(1-e^{-\frac{d\sin |b|}{z_s}}\bigg)
\end{split}
\end{align}

This equation becomes ill-defined for $|b| \ll 1$ when we use a realistic map for $A_V(l,b,\infty)$ and thus we set $b = \text{max}(|b|,1\degree)$ which makes the function well behaved for all angles. We obtain $A_V(l,b,\infty)$ from 2D dust extinction map by \cite{schlegel_maps_1998}, implemented in \texttt{dustmaps} package \citep{green_dustmaps_2018}. To convert $A_V$ to appropriate babd-passes, we use the result from \cite{schlafly_measuring_2011} and get $A_b=[1.37, 1.02, 0.73, 0.54, 0.43, 0.35]\times A_V$ for $u,g,r,i,z,y$ bands, respectively.

\bibliography{references,RefDWDPPS2025}
\bibliographystyle{aasjournalv7}



\end{document}